# Polarization forces in the vicinity of nanoparticles in weakly ionized plasma

M.N. Shneider

*Mechanical and Aerospace Engineering Department, Princeton University, Princeton, New Jersey 08544, USA*
e-mail: m.n.shneider@gmail.com

*It is shown that the polarization forces in a weakly ionized plasma lead to a substantial increase in the fluxes of neutral atoms and molecules to the surface of charged nanoparticles. Thus, the nanoparticles can change thermal balance due to the acceleration of atoms and molecules in the dipole potential and subsequent inelastic collisions to the nanoparticles.*

Carbon and others nanoparticles can be efficiently synthesized in weakly-ionized non-equilibrium plasmas [1,2], equilibrium and non-equilibrium arcs in an inert gas atmosphere [3-5] and in laser ablation plasmas [6,7]. Moreover, it is known that hybrid methods based on chemical vapor deposition (CVD) with a weakly ionized plasma substantially improve the capabilities and characteristics of nanoparticles synthesis [8, 9].

Nanoparticle and soot particulate growth is driven by influxes of plasma ions, neutral atoms and molecules. However, in a weakly ionized plasma the density of neutral particles is many orders of magnitude greater than the density of ions, and their flux to the nanoparticle is much less intense than the flux of neutral particles, even taking into account the Coulomb interaction, which is limited by the Debye screening. Thus, it can be assumed that the growth of carbon nanoparticles and soot particles is mainly due to the influx of carbon atoms and neutral molecules. We will show that in a plasma this influx can increase significantly due to the polarization forces acting on the neutral atoms and molecules in the vicinity of the charged nanoparticles.

The polarization forces have been well characterized and are one of the types of ponderomotive forces [10]. These forces, for example, are responsible for electrostrictive phenomena in liquid and solid dielectrics [10-12], have been successfully applied in optical tweezers [13], and can be used to capture and transport the atoms/molecules of a gas in the optical lattice [14, 15]. Note, that polarization forces, resulting in the formation of chains of clusters or aggregates of clusters, are well studied and known in nanotechnology for a long time (see, for example [16,17]). However, the role of the polarization forces in the transport of neutral atoms and molecules from the surrounding plasma to the nanoparticle, as far as we know, has not been considered.

The chemical composition of the plasma, wherein the synthesis of the nanoparticles takes place, can be very complex (as demonstrated by the results of calculations to determine the plasma composition in an arc between graphite electrodes in a helium atmosphere [18,19]). Let us consider a model non-equilibrium weakly ionized plasma in a mixture of helium atoms, carbon atoms, and molecules. To simplify the analysis, due to a significant difference between the ionization potential (~ 11.3 for carbon versus 24.6 eV for helium), we will consider only atomic carbon ions $C^+$, assuming their density is equal to the electron density of the plasma, $n_i \approx n_e$. At the same time, we assume that the concentration of nanoparticles or soot particles is not very high, so we can suppose that, due to particles charging, the electron density in the plasma varies slightly. The results, presented in this letter are qualitatively the same for different types of ions (e.g., $C^+$ or $C_2^+$) Quantitative changes are also insignificant.

It is known that particles in plasma acquire a charge based on the negative floating potential. The floating potential is on the order of the local electron temperature, $\varphi_s \sim -T_e$, when the absolute value of the electron and ion current equalize [20-22]. Thus, the magnitude of the floating potential $\varphi = \varphi_s$ does not depend on the size of the particles and is determined by the condition $I_{e,tot} + I_i = 0$, where $I_{e,tot}, I_i$ correspond to the total current of electrons on the particle surface and ion current respectively. In general, the thermionic emission and field emission with Schottky effect must be taken into account in the electron current $I_{e,tot}$, however, in this study we do not consider these processes.

Let us consider the spherical nanoparticles in the plasma, for which the conditions of orbit motion limited (OML) approximation are valid, which were developed in the theory of probes and effectively used in dusty plasma theory [20-22]. In this case, the inequality $a \ll \lambda_D \ll l_{i,e}$ is valid Where $a$ is the radius of the nanoparticles; $\lambda_D = \sqrt{\varepsilon_o T_{e[eV]}/en_e}$ is the Debye length; $l_i$, and $l_e$ are mean free paths for ions and electrons, respectively. In this approximation, the effective cross sections for electrons and ions collisions with the negatively charged particle are [20-22]

$$\sigma_e(v) = \begin{cases} \pi a^2 \left(1 - \dfrac{2e|\varphi|}{mv^2}\right), & \dfrac{2e|\varphi|}{mv^2} < 1 \\ 0, & \dfrac{2e|\varphi|}{mv^2} > 1 \end{cases} \qquad (1)$$

$$\sigma_i(v) = \pi a^2 \left(1 + \dfrac{2e|\varphi|}{Mv^2}\right) \qquad (2)$$

Assuming that in a quasi-neutral plasma the electrons and ions have Maxwellian distribution functions, corresponding to $T_e$ и $T_i$ temperatures, respectively, and by equating the electron and ion currents from the plasma to the surface of the particle, we obtain the equation for the floating potential acquired by the particle [21]

$$|\varphi_s| \approx \dfrac{kT_e}{e} \ln\left[\sqrt{\dfrac{MT_e}{mT_i}} \Big/ \left(1 + \dfrac{e|\varphi_s|}{kT_i}\right)\right], \qquad (3)$$

where $m, M$ are electron and ion masses.

As an example, Fig. 1 shows the electron temperature dependence of the floating potential acquired by spherical nanoparticle in a weakly ionized nonequilibrium plasma, under the assumptions that $C^+$ is the primary ion and the ion temperature $T_i = 300\,\text{K}$.

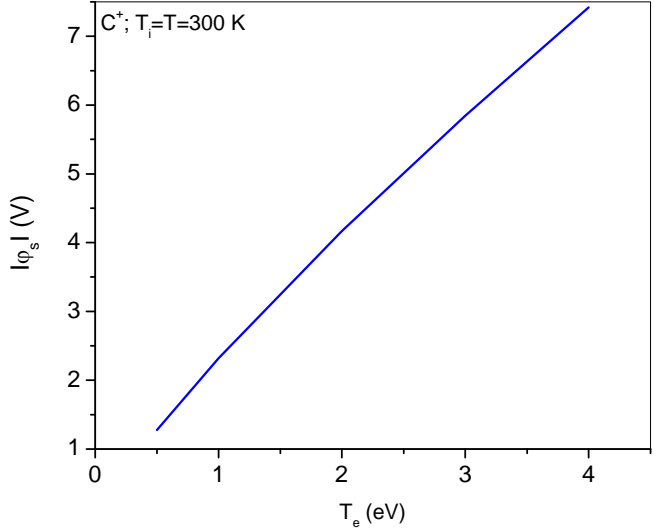

**Fig. 1.** Floating potential acquired by spherical nanoparticles in the plasma as a function of electron temperature. It is assumed that the main ion $C^+$ and the ion temperature $T_i = 300\,\text{K}$.

It should be noted that, in general, nanoparticles can hold a limited number of electrons, referred as their "charge limit" due to the competition between electron affinity and inter-electron Coulomb repulsion. Here, we will neglect with this effect, assuming that the floating potential is below the affinity energy for electrons attached to the surface [23].

The electrostatic field in the vicinity of a charged spherical particle of radius $a$, which is under a floating potential is

$$E(r) = \frac{\varphi_s a}{r^2}, \quad r \geq a. \tag{4}$$

At the typical values of a floating potential shown in Fig. 1 for nanoparticles of small size ($a \sim 1-5\,\text{nm}$), the field (4) can be quite large ($\sim 10^9$ V/m), which is significantly higher than the field in the Debye layer (ambipolar field), $E_D \sim T_e / \lambda_D$. For example, for typical conditions of the arc discharge, in which the synthesis of nanoparticles occurs, in the axial region $T_e \sim 1\,\text{eV}$ and $n_e \sim 10^{22}\,\text{m}^{-3}$ [24], we estimate the maximum ambipolar field $E_D \sim 1.35 \cdot 10^7$ V/m. Note, that the Stark broadening in a plasma in the presence of nanoparticles, which are charging to the floating potential, should significantly exceed the Stark broadening determined by the ambipolar field. Such a sharp increase in the Stark broadening can serve as additional evidence of the initial stage of the synthesis of nanoparticles in a weakly ionized plasma.

It is known that in an electric field any atoms and non-polar molecules become polarized and attain an induced dipole moment. Polar molecules have their own permanent dipole moment (for example, water molecules) and try to orient along the electric field and obtain an additional dipole moment due to polarization [10,11]. The dipole moment in the electric field for atoms and non-polar molecules is $\vec{p}_d = \alpha \vec{E}$, and $\vec{p}_d = \alpha_{eff} \vec{E}$ for polar molecules. Here $\alpha$ is the electronic polarizability of the particle; $\alpha_{eff} = \alpha + \frac{\mu_{d0}^2}{3kT}$ is the effective polarizability of polar molecules with permanent dipole moment $\mu_{d0}$. Electronic polarizability depends on the electric field frequency and, in general, is a tensor for molecules. We will consider the atoms and nonpolar symmetrical molecules in quasi-static electric field, for which we can operate with average values of the electronic polarizability. For example, the values of the polarizabilities for helium and carbon atoms and molecules C, $C_2$, $C_3$, and $C_4$ in SI units are, respectively: $\alpha = 2.22 \cdot 10^{-41}$, $1.95 \cdot 10^{-40}$, $3.5 \cdot 10^{-40}$, $5.36 \cdot 10^{-40}$, and $8.22 \cdot 10^{-40}$ $C^2 m^2 J^{-1}$ [19].

The force acts on induced dipoles in the direction of the maximum electric field, regardless of the sign of the field [10, 11]

$$\vec{F}_d = -\nabla U_d, \tag{5}$$

where $U_d = -\frac{1}{2}\alpha E^2$ is the additional potential energy of each atom or molecule in an electric field (a characteristic dipole potential). Taking into account (4), the dipole potential of atoms and nonpolar molecules in the vicinity of the charged spherical nanoparticles is

$$U_d = -\frac{1}{2}\alpha E^2 = -\frac{\alpha}{2}\left(\frac{\varphi_s a}{r^2}\right)^2, \quad r \geq a. \tag{6}$$

If the value of the dipole potential (6) in weakly ionized plasma is greater than or comparable to the energy of the thermal motion of the atoms/molecules of the background weakly-ionized plasma, the motion of the neutral gas particles will be significantly influenced by force (5).

The dependency of the dipole potential of carbon atoms and molecules in the vicinity of the surface of spherical nanoparticles charged to the floating potential, on the size of the nanoparticles is illustrated in Fig. 2. The dipole potential is shown in units of $kT$.

Fig. 2a shows the values $|U_d|/kT$ for carbon atoms and the various values of the electron temperature in plasma, and Fig 2b the dipole potential for various atoms and molecules of carbon, at chosen electron temperatures, for example, the electron temperature $T_e = 2$ eV. The temperatures of the background neutral gas particles and ions were taken to be $T = T_i = 300$ K.

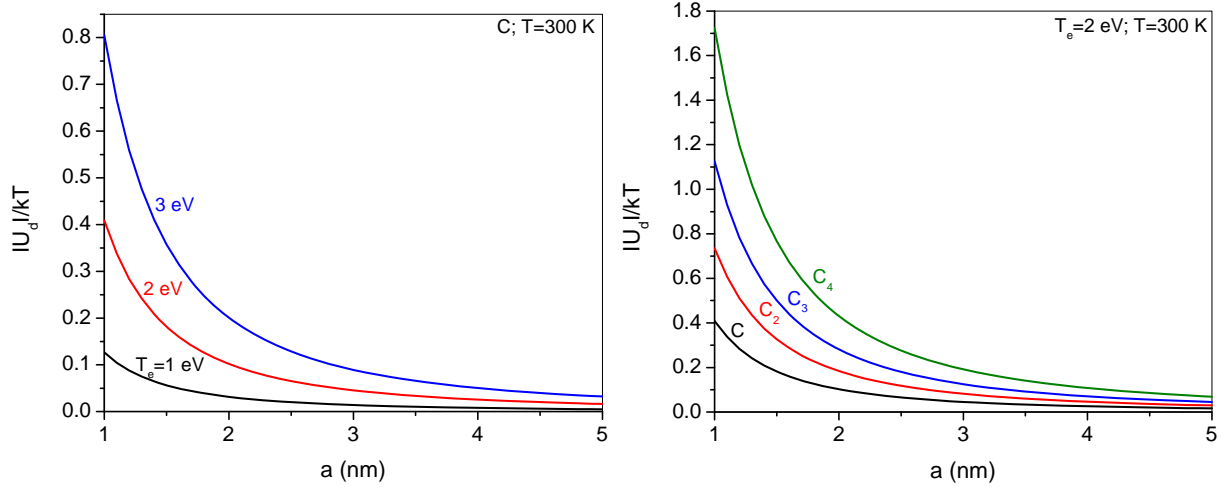

Fig.2. Dipole potential in units $kT$ depending on the radius of the spherical nanoparticles when the primary ion $C^+$, and $T = T_i = 300$ K. (a) - for atoms $C$ at various electron temperatures $T_e$ in plasma; (b) - for carbon atoms and different molecules at assumed $T_e = 2$ eV in plasma.

We see that for the charged nanoparticles of size $a \sim 1$ nm their impact on the neutral component of the background weakly-ionized plasma can be substantial. This leads to an important conclusion: the flow of neutral atoms and molecules to the nanoparticles in plasma can be substantially higher than in the case of a neutral nanoparticles.

In the Knudsen regime when the mean free paths of atoms and molecules are greater than the size of the nanoparticles the motion of neutral gas particles in the vicinity of the charged nanoparticles can be described by OLM approximation. For example, the OLM approximation could be used to describe neutral particle motion in rather weakly ionized plasma when the field of the charged nanoparticles is not screened by plasma, i.e. when the condition $\lambda_D > l_n \gg a$ holds. For spherical nanoparticles of size $a < 10$ nm this is true up to atmospheric pressure, at gas temperature $T \geq 300$ K, electron temperature $T_e \sim 1$ eV and unperturbed density of electrons and ions in the plasma $n_e \approx n_+ < 10^{22}$ m$^{-3}$. In this case, the scattering of atoms/molecules on the nanoparticle is determined by the effective cross-section . The effective cross-section is similar to the cross section for the scattering of positive ions on the negatively-charged nanoparticles (2), with the corresponding value of dipole potential $U_d$ (6) at $r = a$ used instead of the electrostatic potential energy for a singly charged ion $e\varphi_s$:

$$\sigma_d(v) = \pi a^2 \left(1 + \frac{2|U_d(a)|}{Mv^2}\right). \tag{7}$$

Assuming a Maxwell distribution function of the atoms/molecules in the plasma

$$f(v) = \left(\frac{M}{2\pi k_B T}\right)^{3/2} \exp\left(-\frac{Mv^2}{2k_B T}\right), \tag{8}$$

and using a cross-section (7), we find that the number of atoms, colliding with the surface of a charged spherical nanoparticles per unit time:

$$4\pi a^2 \Gamma_d = 4\pi N \int_0^\infty v^3 \sigma_d(v) f(v) dv = 4\pi a^2 \Gamma_0 \left(1 + |U_d(a)|/kT\right). \tag{9}$$

Here, $\Gamma_0 = \frac{1}{4} N \bar{v}$, $\Gamma_d$ are fluxes of atoms/molecules to the surface of nanoparticle with and without taking into account the polarization forces, respectively; $N$ is the density of neutral atoms or molecules in the ambient plasma; $\bar{v} = \left(\frac{8kT}{\pi M}\right)^{1/2}$ is the average thermal velocity of the gas particles. It follows from (9)

$$\Gamma_d / \Gamma_0 = \left(1 + |U_d(a)|/kT\right). \tag{10}$$

As we can see from Fig. 2, where values $|U_d|/kT$ are shown, dipole forces can lead to a noticeable increase in the flux of neutral particles from the plasma to the charged nanoparticles. Therefore, the growth of nanoparticles or soot particles in non-equilibrium weakly-ionized plasma can occur much faster than in non-ionized gas, especially at the initial stage of the evolution of nanoparticles when their size is relatively small. All this is even more noticeable for the neutral clusters, which have a much greater polarizability than atoms or simple molecules. For spherical particles larger than $a \sim 5$ nm, the effects associated with polarization forces become insignificant. It is a different matter however, if the surface of the nanoparticle has large inhomogeneities in which the field is enhanced markedly. In this case, the non-uniform protrusions will grow faster than the rest of the surface areas of large nanoparticles due to the increased flux of neutral atoms and molecules from the plasma to the areas with the sharply increased electric field. This can lead to the emergence and growth of whiskers in plasma [16], and a pronounced fractal structure of soot particles [17].

Accounting for the dipole potential can radically change the heat balance of nanoparticles. The neutral atoms and molecules, colliding with and captured by the nanoparticles, transfer to it an averaged energy $\sim \frac{3}{2} kT + |U_d|$ in each collision event. as they acquire additional energy due to the work of the dipole force, which accelerates polarized gas particles towards the charged nanoparticles. As a result, there is another reason for the higher temperature of nanoparticles than the gas particles temperature in the ambient plasma, in addition to those considered in [25-27]. In the case of one type of neutral atoms or molecules, an estimate obtained in OML approximation for the additional heat source:

$$Q_d = 4\pi N \int_0^\infty \left(\frac{Mv^2}{2} + |U_d(a)|\right) v^3 \sigma_d(v) f(v) dv = \frac{4}{3} \pi N a^2 \bar{v} \bar{\varepsilon} \left[1 + \frac{|U_d(a)|}{kT} + \frac{1}{2}\left(\frac{|U_d(a)|}{kT}\right)^2\right], \tag{11}$$

where $\bar{\varepsilon} = \frac{3}{2}kT$ is the average energy of the translational motion of neutral atoms/molecules. If there are different atoms and molecules deposited on the growing nanoparticle (e.g., C, $C_2$, $C_3$, ... etc), the total of their contribution to the heating of nanoparticles is

$$Q_d = \frac{4}{3}\pi a^2 \bar{\varepsilon} \sum_i N_i \bar{v}_i \left[1 + \frac{|U_{d,i}(a)|}{kT} + \frac{1}{2}\left(\frac{|U_{d,i}(a)|}{kT}\right)^2\right], \quad (12)$$

where $N_i, \bar{v}_i$, and $U_{d,i}$ are the density, the average thermal velocity and the dipole potential of the $i$-th neutral component. As we can see, at $|U_{d,i}(a)|/kT > 1$, the contribution of the polarization forces, defined by the terms, which depend on the dipole potential in the formulas (11) and (12), can be quite substantial.

**Conclusions**

We have shown that the polarization forces in a weakly ionized plasma, acting on neutral atoms and molecules, lead to a substantial increase in their fluxes to the charged nanoparticles and changes in the heat balance. All this may have a noticeable impact on the dynamics of nanoparticle synthesis and growth of soot particles.

I would like to thank Drs. Y. Raitses, I. Kaganovich, V. Nemchinskiy, and Mr. J. Mitrani for useful discussions. This work was supported by the U.S. Department of Energy, Office of Science, Basic Energy Sciences, Materials Sciences and Engineering Division.